\documentclass[journal]{IEEEtran}
\usepackage{placeins}
\usepackage{flafter}
\usepackage{diagbox}
\usepackage[inline]{enumitem}
\usepackage{xcolor}
\usepackage{amsmath,amssymb}
\usepackage{algorithm}
\usepackage[group-separator={,},group-minimum-digits=4]{siunitx}
\usepackage{threeparttable}
\usepackage{array}
\usepackage[caption=false,font=normalsize,labelfont=sf,textfont=sf]{subfig}
\usepackage{textcomp}
\usepackage{stfloats}
\usepackage{url}
\usepackage{verbatim}
\usepackage{graphicx}
\usepackage{algpseudocode}
\usepackage{cite}
\usepackage{balance}
\usepackage{mathrsfs}
\usepackage[colorlinks=true,
            linkcolor=blue,
            anchorcolor=blue,
            citecolor=blue]{hyperref}

\makeatletter
\def\@IEEENORMtitlevspace{1.2\baselineskip}
\def\@IEEEMINtitlevspace{1.0\baselineskip}
\g@addto@macro\normalsize{%
  \setlength{\abovedisplayskip}{3.5pt plus 1pt minus 1pt}%
  \setlength{\belowdisplayskip}{3.5pt plus 1pt minus 1pt}%
  \setlength{\abovedisplayshortskip}{0pt plus 1pt}%
  \setlength{\belowdisplayshortskip}{3.5pt plus 1pt minus 1pt}}
\def\section{\@startsection{section}{1}{\z@}{1.8ex plus 0.5ex minus 0.3ex}%
  {0.7ex plus 0.3ex}{\normalfont\normalsize\centering\scshape}}
\def\subsection{\@startsection{subsection}{2}{\z@}{2.0ex plus 0.5ex minus 0.3ex}%
  {0.7ex plus 0.3ex}{\normalfont\normalsize\itshape}}
\makeatother

\begin{document}

\title{ 
\fontsize{16}{19.2}\selectfont
{Fading Suppression in Distributed Acoustic Sensing Using Frequency-Domain Replica Reconstruction and Noise-Whitened Principal Component Combining}  
}  
\author{ 
{
Ziang~Chen, Zhiyang~Xue, Huan~Huang,~\textit{Member,~IEEE}, Yuhan~Jiang, Dongdong~Zou, Gangxiang~Shen,~\textit{Senior~Member,~IEEE,~Fellow,~Optica}, and Yi~Cai,~\textit{Fellow,~Optica}
}
\thanks{  

Z.~Chen, Z.~Xue, H.~Huang, Y.~Jiang, D.~Zou, G.~Shen, and Y.~Cai are with the School of Electronic and Information Engineering, Soochow University, Suzhou, Jiangsu 215006, China 
(e-mail:  \nolinkurl{zachen@stu.suda.edu.cn}; \nolinkurl{zyxue999@stu.suda.edu.cn}; \nolinkurl{hhuang1799@gmail.com}; \nolinkurl{yuhanjiangsuda@outlook.com}; \nolinkurl{ddzou@suda.edu.cn};  \nolinkurl{shengx@suda.edu.cn}; \nolinkurl{yicai@suda.edu.cn}).  
}
}
\maketitle

\begin{abstract}
Frequency-selective Rayleigh fading in distributed acoustic sensing (DAS) based on phase-sensitive optical time-domain reflectometry (\(\phi\)-OTDR) calls for effective diversity, but existing rotated-vector-sum (RVS) combining of nonoverlapping subbands incurs a spatial-resolution penalty. In this letter, we propose frequency-domain replica reconstruction and noise-whitened principal component combining (PCC) that combines a full-band replica and overlapping subbands using fixed weights estimated across probing periods after whitening correlated replica noise. This approach retains wider subbands and reduces reliance on noisy first-period alignment, balancing diversity with nominal spatial resolution. Experiments over approximately \(7.1\)~km of fiber with \(16~\mu\mathrm{s}\) linear frequency-modulated (LFM) pulses recover \(500\)-Hz vibrations with reduced background phase excursions and a \(2.46\)--\(2.76\)-dB improvement in the vibration-to-background ratio (VBR) over the first-period RVS. 
\end{abstract}


\section{Introduction}\label{Intro}
\IEEEPARstart{D}{istributed} acoustic sensing (DAS) provides spatially resolved vibration measurements over long fiber links, enabling dense seismic observations using optical fibers~\cite{JoussetNC2018}. Phase-sensitive optical time-domain reflectometry (\(\phi\)-OTDR) implements DAS by probing coherent Rayleigh backscattering. With rectangular pulses, finer spatial resolution requires shorter pulses, but reduces launched energy at fixed peak power. Maintaining pulse energy then requires higher peak power, but peak power is limited by fiber nonlinearities~\cite{MartinsOL2013}. Optical pulse compression relaxes this tradeoff. Long linear frequency-modulated (LFM) pulses increase energy at fixed peak power, while matched filtering makes the nominal spatial resolution depend on the sweep bandwidth~\cite{ZouOE2015,LuOL2017}.

Destructive interference among Rayleigh backscatters produces deep fades that impair phase recovery~\cite{LiokumovichJLT2015}. Their frequency dependence allows sufficiently-separated probe frequencies to provide decorrelated responses~\cite{ZhouJLT2013}. Frequency and wavelength diversity~\cite{GuPJ2020} and spatial diversity with few-mode~\cite{ZhaoOE2021} or multicore fibers~\cite{XiaoOE2024} provide multiple observations of the same perturbation. However, these schemes consume additional spectral or spatial resources and may require extra optical sources, multiplexers, or receiver channels.

Diversity can also be extracted digitally from a single probe return. For rectangular pulses, spectrum extraction and remix combines partly overlapping segments of the sinc-shaped spectrum~\cite{WuJLT2019}. The trace-decomposition layer of three-layer structure multiplexing similarly combines the main lobe and two spectral sidelobes~\cite{WuJLT2024}. However, for LFM pulses with large time-bandwidth products, energy is concentrated within the swept band, yielding weak out-of-band replicas. To this end, inner-pulse frequency division instead partitions the chirped return into nonoverlapping subbands and coherently combines their responses using rotated-vector-sum (RVS) combining~\cite{ChenOE2017}. However, equally partitioning a fixed sweep bandwidth reduces each subband's bandwidth and increases its nominal spatial-resolution scale in proportion to the number of branches, trading spatial resolution for diversity.

In this work, we propose a fading suppression scheme based on frequency-domain replica reconstruction and noise-whitened principal component combining (PCC). Frequency-domain filtering reconstructs a full-band replica and overlapping subband replicas from each LFM return, retaining wider subbands than nonoverlapping partitioning.  
The proposed PCC forms an uncentered correlation matrix from noise-whitened replicas across probing periods and derives fixed combining weights from its dominant eigenvector, reducing reliance on noisy first-period alignment. 
Experiments over about \(7.1\)~km of fiber demonstrate \(500\)-Hz vibration recovery with improved background rejection compared to the first-period RVS.

\section{Signal Model and Subband Diversity}\label{Principle}
\subsection{LFM Probing and Matched Filtering}\label{SysMod}

\begin{figure}[!t]
    \centering
    \includegraphics[scale=0.75]{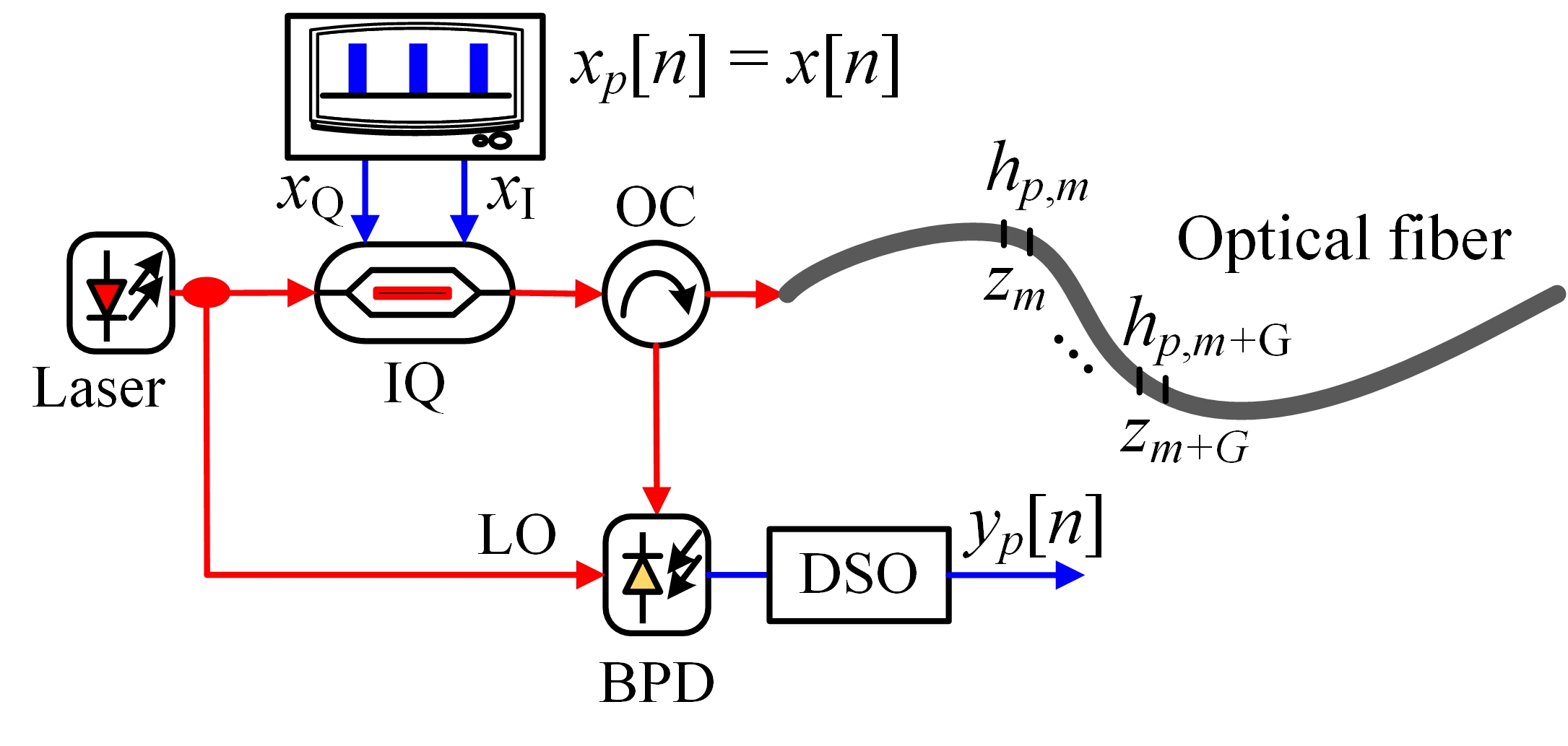}
    \caption{Schematic of the coherent \(\phi\)-OTDR system with heterodyne reception. IQ: in-phase/quadrature modulator; OC: optical circulator; BPD: balanced photodetector; DSO: digital storage oscilloscope.} 
    \label{fig1}
\end{figure}

Fig.~\ref{fig1} shows a \(\phi\)-OTDR system in which a narrow-linewidth laser supplies the probe and local oscillator (LO). The normalized envelope of the \(p\)-th LFM pulse, \(p=1,\ldots,P\), is
\begin{equation}
    x_p(t)=a(t)\exp\!\left[j\pi K\left(t-\frac{T_{\mathrm b}}{2}\right)^2\right],
    \label{eq:lfm_probe}
\end{equation}
where \(a(t)=1\) for \(0\le t<T_{\mathrm b}\) and zero otherwise, \(T_{\mathrm b}\) is the pulse duration, and \(B\) is the sweep bandwidth. The chirp rate \(K=B/T_{\mathrm b}\) gives a sweep over \([-B/2,B/2]\)~\cite{ZouOE2015,LuOL2017}. For absolute time \(u\) and pulse repetition interval \(T_{\mathrm r}\), the probe and LO fields are proportional to \(x_p(u-pT_{\mathrm r})e^{j2\pi\nu_{\mathrm c}u}\) and \(e^{j2\pi\nu_{\mathrm{LO}}u}\), where \(\nu_{\mathrm c}\) and \(\nu_{\mathrm{LO}}\) are their carrier frequencies. Digital upconversion to \(f_{\mathrm{IF}}=\nu_{\mathrm c}-\nu_{\mathrm{LO}}\), with \(|f_{\mathrm{IF}}|>B/2\), provides the heterodyne offset. Then, the upconverted waveform drive the in-phase/quadrature (IQ) modulator.

Approximate the complex baseband as bandlimited to \(B\), and use equivalent Nyquist sampling with \(F_s=B\) and \(T_s=1/B\). A \(L\)-length fiber is divided into \(M=\lceil L/\Delta z\rceil\) bins at \(z_m=m\Delta z\), \(m=0,\ldots,M-1\), where \(\Delta z=v_{\mathrm g}T_s/2=v_{\mathrm g}/(2B)\), \(\tau_m=2z_m/v_{\mathrm g}=mT_s\), and \(v_{\mathrm g}\) is the group velocity. This grid spacing is equal to the nominal full-band resolution scale. 
Under conditions of weak scattering and approximately uniform phase perturbation within each bin, the equivalent Rayleigh-backscatter coefficient is~\cite{LiokumovichJLT2015}
\begin{equation}
    h_{p,m}
    =\underbrace{\rho_m e^{-2\alpha z_m}e^{-j2\beta_{\mathrm c}z_m}}_{c_m}
    e^{j\phi_p(z_m)},
    \label{eq:rayleigh_channel}
\end{equation}
where \(\rho_m\) aggregates microscopic scattering, intrabin static phases, and a fixed polarization projection, \(\alpha\) is the one-way field attenuation coefficient, and \(\beta_{\mathrm c}\) is the propagation constant at \(\nu_{\mathrm c}\). Many randomly phased contributions make \(\rho_m\) approximately circularly symmetric complex Gaussian, with Rayleigh-distributed magnitude. Neglecting background drift and residual laser phase noise, \(c_m\) remains fixed over the observation interval, while \(\phi_p(z_m)\) is the vibration-induced round-trip phase change from an unperturbed state~\cite{ChenOE2017}. At this unperturbed state, \(\phi_p(z_m)=0\) and \(h_{p,m}=c_m\), with a generally nonzero static phase \(\arg c_m\). The first probe (\(p=1\)) need not correspond to this state.

Assume nonoverlapping returns (\(T_{\mathrm r}\ge T_{\mathrm b}+2L/v_{\mathrm g}\)), a flat receiver response, and negligible phase evolution within each return, while allowing \(\phi_p(z_m)\) to vary across probing periods. Balanced detection and digital quadrature downconversion generate a baseband return~\cite{ZouOE2015,LuOL2017}. Sampling at \(t=u-pT_{\mathrm r}=nT_s\) yields the finite-memory linear convolution~\cite{HuangDMTarXiv2026,HuangCPOFDMarXiv2026}
\begin{equation}
    y_p[n]=\sum_{m=0}^{M-1}h_{p,m}x_p[n-m]+w_p[n],
    \label{eq:discrete_received}
\end{equation}
where \(x_p[n]=x_p(nT_s)\) and \(w_p[n]\) is the receiver noise.

For repeated probes \(x_p[n]=x[n]\) with energy \(E_x=\sum_n|x[n]|^2\), matched filtering gives~\cite{ZouOE2015,LuOL2017}
\begin{equation}
    \widehat h_{p,m}
    =\frac{1}{E_x}\sum_n y_p[n]x^*[n-m]
    =\sum_{\ell=0}^{M-1}h_{p,\ell}r_x[m-\ell]+\eta_{p,m},
    \label{eq:mf_channel}
\end{equation}
where \(r_x[q]=E_x^{-1}\sum_n x[n]x^*[n-q]\) is the normalized autocorrelation, \(r_x[0]=1\), and \(\eta_{p,m}\) denotes the filtered noise. Nonzero-lag correlation introduces leakage from other range bins, i.e.,  spatial inter-symbol interference (ISI)~\cite{HuangDMTarXiv2026,HuangCPOFDMarXiv2026}.

\subsection{Subband Diversity Using First-Period RVS}\label{OERVS}
Inner-pulse frequency division mitigates fading by partitioning the probe into \(V\) nonoverlapping subbands of width \(B_{\mathrm s}=B/V\)~\cite{ChenOE2017}. Let \(x^{(v)}[n]\), \(v=1,\ldots,V\), be the ideally bandpass-filtered probe retaining its original time origin, with energy \(E_x^{(v)}=\sum_n|x^{(v)}[n]|^2\). Replacing the template and energy in~\eqref{eq:mf_channel} by \(x^{(v)}\) and \(E_x^{(v)}\) gives
\begin{equation}
    \widehat h^{(v)}_{p,m}
    \simeq c_m^{(v)}e^{j\phi_p(z_m)}+\eta^{(v)}_{p,m},
    \label{eq:subband_channel}
\end{equation}
where \(c_m^{(v)}=\sum_{\ell=0}^{M-1}c_\ell r_x^{(v)}[m-\ell]\). 
The common-phase approximation in~\eqref{eq:subband_channel} requires nearly-constant \(\phi_p(z)\) over the subband's spatial response~\cite{ChenOE2017}. Locally homogeneous, spatially uncorrelated Gaussian scattering results in approximately independent subband fading. Although the nominal spatial resolution of each subband degrades to \(v_{\mathrm g}/(2B_{\mathrm s})=V\Delta z\), all branches retain the grid \(z_m\).

First-period (\(p=1\)) phase alignment gives RVS~\cite{ChenOE2017}
\begin{equation}
    \begin{aligned}
        \overline h_{p,m}
        &=\frac{1}{V}\sum_{v=1}^{V}\widehat h^{(v)}_{p,m}
        e^{-j\arg\widehat h^{(v)}_{1,m}}\\
        &\simeq C_m e^{j[\phi_p(z_m)-\phi_1(z_m)]}
        +\overline\eta_{p,m},
    \end{aligned}
    \label{eq:first_frame_rvs}
\end{equation}
where \(C_m=V^{-1}\sum_v|c_m^{(v)}|e^{-j\epsilon_m^{(v)}}\) and \(\overline\eta_{p,m}\) is the rotated and combined noise. For nonzero \(c_m^{(v)}\) and reference responses, the first-period noise \(\eta^{(v)}_{1,m}=(E_x^{(v)})^{-1}\sum_n w_1[n]x^{(v)*}[n-m]\) induces the wrapped alignment error
\begin{equation}
    \epsilon_m^{(v)}
    =\arg\!\left(1+\frac{\eta^{(v)}_{1,m}}{c_m^{(v)}e^{j\phi_1(z_m)}}\right).
    \label{eq:reference_phase_error}
\end{equation}

With error-free alignment, \(C_m>0\) and the fields add constructively. For gauge length \(L_{\mathrm g}=G\Delta z\), \(1\le G<M\), define \(g_{p,m}=\overline h_{p,m+G}\overline h_{p,m}^{\,*}\), \(m=0,\ldots,M-G-1\). The recovered gauge-differential phase is $\Delta\widehat\phi_{p,m}
    =\operatorname{unwrap}_p\!\left\{\arg g_{p,m}\right\}$,
where \(\operatorname{unwrap}_p\) denotes slow-time unwrapping over reliable continuous records. Since \(\overline h_{1,m}=V^{-1}\sum_v|\widehat h^{(v)}_{1,m}| \), the first phase is zero. For \(p>1\), negligible current-probe noise and correct unwrapping give the true gauge-differential phase change \([\phi_p(z_{m+G})-\phi_p(z_m)]-[\phi_1(z_{m+G})-\phi_1(z_m)]\) plus \(\arg C_{m+G}-\arg C_m\). However, zeroing the first phase cannot remove the first-period noise \(\eta^{(v)}_{1,m}\), which also reduces \(|C_m|\) and increases sensitivity to subsequent noise.

\begin{figure*}[!b]
    \centering
    \includegraphics[width=0.88\textwidth]{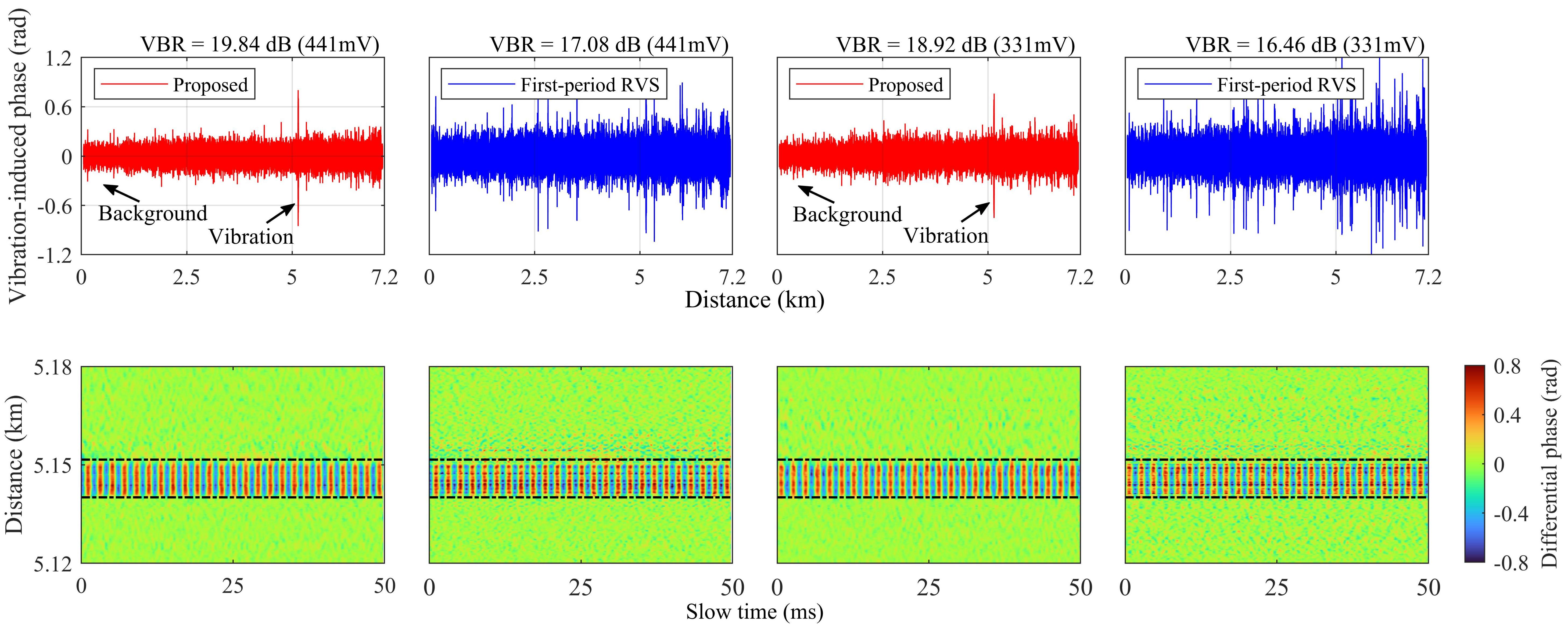}
    \caption{Experimental vibration-induced phase for the proposed principal component combining and first-period rotated-vector-sum (RVS) combining at \(441\) and \(331\)~mV$_{\mathrm{pp}}$ AWG settings, where the upper row overlays phase traces from all probing periods with vibration-to-background ratio (VBR) defined as \(20\log_{10}(\mu_{\mathrm v}/\mu_{\mathrm b})\) using mean absolute phase over vibration observations inside (\(\mu_{\mathrm v}\)) and outside (\(\mu_{\mathrm b}\)) the vibration observation window, and the lower row shows local slow-time--distance maps.}
    \label{fig:recovered_phase}
\end{figure*}

\section{Frequency-Domain Replica Reconstruction and Noise-Whitened PCC}\label{Proposed}

\begin{figure*}[!t]
    \centering
    \includegraphics[width=0.88\textwidth]{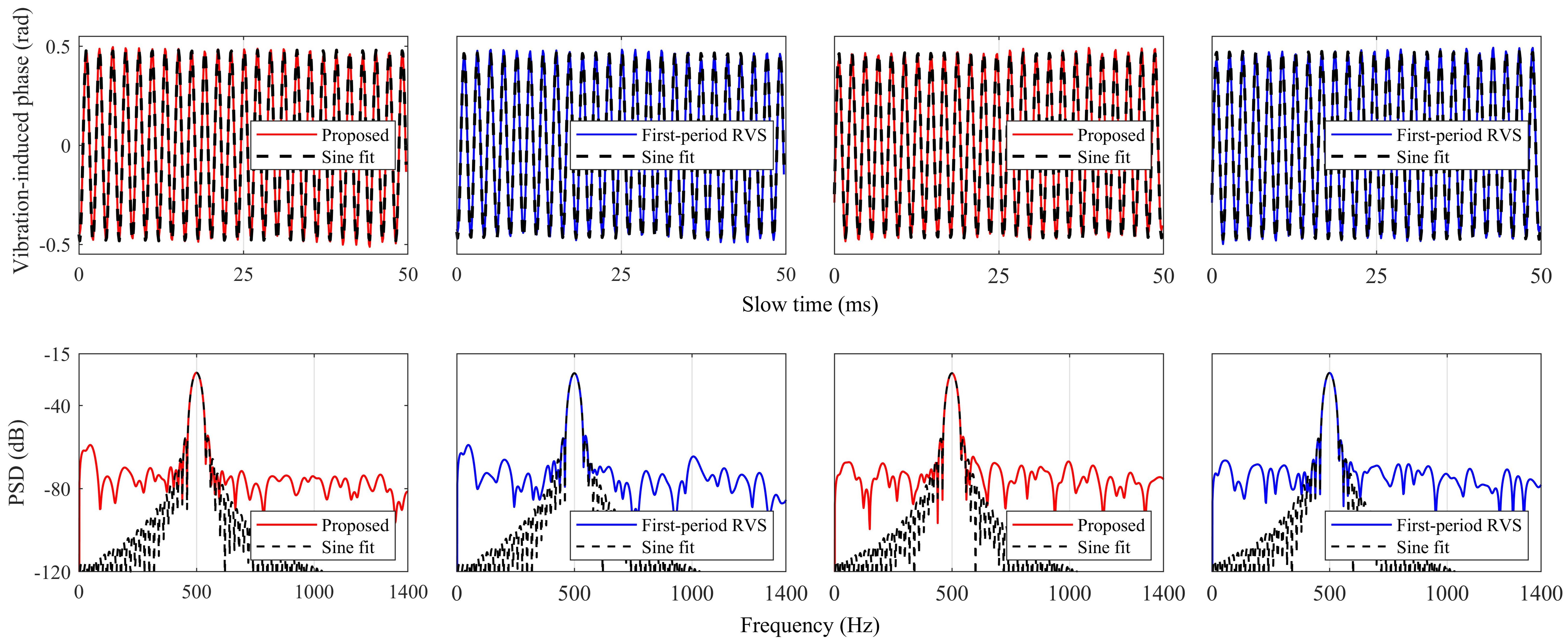}
    \caption{Recovered vibration-induced phase and power spectral density (PSD), where the upper row compares measured phase traces (solid) with sine fits (black dashed) and the lower row shows their one-sided PSDs in dB relative to \(1~\mathrm{rad}^2/\mathrm{Hz}\).}
    \label{fig:recovered_phasetrace}
\end{figure*}
To reduce the \(V\)-fold nominal spatial-resolution penalty of nonoverlapping subbands, we reconstruct a full-band replica (\(v=0\)) and \(V\ge2\) overlapping subband replicas. For uniform coverage with adjacent overlap ratio \(0\le\zeta<1\), the subband width is given by
\begin{equation}
    B_{\mathrm o}=\frac{B}{1+(V-1)(1-\zeta)},
    \label{eq:overlap_geometry}
\end{equation}
with centers \(f_v=-B/2+B_{\mathrm o}/2+(v-1)\delta f\), $\delta f=(1-\zeta)B_{\mathrm o}$, \(v=1,\ldots,V\). 

Let \(X[k]\) and \(Y_p[k]\), \(k=0,\ldots,N-1\), be the \(N\)-point discrete Fourier transforms (DFTs) of \(x[n]\) and \(y_p[n]\), sufficiently zero-padded for linear correlation. The replicas are
\begin{equation}
        \widetilde h^{(v)}_{p,m}
        =\frac{1}{N}\sum_{k=0}^{N-1}D_v[k]Y_p[k]e^{j2\pi km/N},
    \label{eq:replica_reconstruction}
\end{equation}
where $D_v[k]=\frac{X^*[k]W_v[k]}{E_x}$, \(W_0[k]=1\) over the probe band, and \(W_v[k]\) is a Hann window of width \(B_{\mathrm o}\) centered at \(f_v\) for \(v\ge1\). All replicas share normalization \(E_x\) and grid \(z_m\). 
Setting \(\zeta=0\) and excluding the full-band replica before calibration gives a bank of nonoverlapping subband replicas. 

To reduce reliance on noisy first-period alignment, PCC~\cite{LiuTWC2007} estimates a combining direction from a calibration set \(\mathcal P_{\mathrm c}\subseteq\{1,\ldots,P\}\), with \(1\le P_{\mathrm c}=|\mathcal P_{\mathrm c}|\le P\). Stack the \(J\) retained replicas (\(J=V+1\) for the full bank) into \(\widetilde{\mathbf h}_{p,m}\). Applying the common-phase approximation in~\eqref{eq:subband_channel} gives 
\begin{equation}
    \widetilde{\mathbf h}_{p,m}
    \simeq \widetilde{\mathbf c}_m e^{j\phi_p(z_m)}
    +\widetilde{\boldsymbol\eta}_{p,m},
    \label{eq:replica_rank_one}
\end{equation}
where \(\widetilde{\mathbf c}_m\) stacks the static responses obtained by filtering \(\{c_\ell\}\) with the inverse DFT of \(D_v[k]X[k]\), and \(\widetilde{\boldsymbol\eta}_{p,m}\) is the reconstructed noise. For \(\widetilde{\mathbf c}_m\ne\mathbf0\), the noiseless replica matrix has rank one. In this case,  every signal outer product equals \(\widetilde{\mathbf c}_m\widetilde{\mathbf c}_m^{\mathrm H}\), independently of \(\phi_p(z_m)\).

For spectrally white receiver noise \(w_p[n]\), the replica-noise covariance is proportional to the Gram matrix
\begin{equation}
    [\boldsymbol\Gamma]_{v,v'}
    =\sum_{k=0}^{N-1}D_v[k]D_{v'}^*[k],
    \label{eq:replica_noise_gram}
\end{equation}
where \(v,v'\) index the retained replicas. Spectral overlap in~\eqref{eq:replica_reconstruction} induces inter-replica correlation in both fading and noise. For regularized noise whitening, define \(\mathbf R_0=J\boldsymbol\Gamma/\operatorname{tr}(\boldsymbol\Gamma)\), with unit average diagonal, and choose \(\mathbf Q\) satisfying
\begin{equation}
    \mathbf Q(\mathbf R_0+\lambda\mathbf I)\mathbf Q^{\mathrm H}=\mathbf I,
    \label{eq:whitenin}
\end{equation}
where \(\lambda>0\) regularizes the noise model, providing approximate whitening. The uncentered calibration correlation matrix is given by
\begin{equation}
    \widehat{\mathbf R}_m
    =\frac{1}{P_{\mathrm c}}\sum_{p\in\mathcal P_{\mathrm c}}
    \mathbf Q\widetilde{\mathbf h}_{p,m}
    \widetilde{\mathbf h}_{p,m}^{\mathrm H}\mathbf Q^{\mathrm H}.
    \label{eq:pcc_combining}
\end{equation}
The dominant unit-norm eigenvector \(\mathbf e_m\) of \(\widehat{\mathbf R}_m\) defines the combining direction in whitened coordinates. The corresponding combining vector in the original replica coordinates is
\begin{equation}
    \mathbf b_m
    =\frac{\mathbf Q^{\mathrm H}\mathbf e_m}{\|\mathbf Q^{\mathrm H}\mathbf e_m\|_1},
    \label{weightvec}
\end{equation}
where \(\mathbf b_m\) stacks \(b_{m,v}\) and \(\|\cdot\|_1\) denotes the \(\ell_1\)-norm; the positive normalization changes only the amplitude scale. For uncorrelated and equal-variance replica noise, setting \(\mathbf Q=\mathbf I\) in~\eqref{eq:pcc_combining}--\eqref{weightvec} gives \(\mathbf b_m=\mathbf e_m/\|\mathbf e_m\|_1\).

Pooling calibration outer products reduces sampling fluctuations for zero-mean, temporally independent noise uncorrelated with the signal. However, combining-direction estimation still depends on the accuracy of the noise model. The calibrated weights remain fixed and combine each probe as \(h^{\mathrm{PCC}}_{p,m}=\mathbf b_m^{\mathrm H}\widetilde{\mathbf h}_{p,m}\), preserving \(\phi_p(z_m)\) up to a static offset without averaging the recovered phase across probing periods. Gauge differencing then yields
\begin{equation}
        \Delta\widehat\phi^{\mathrm{PCC}}_{p,m}
        =\operatorname{unwrap}_p\!\left\{
        \arg\!\left[g^{\mathrm{PCC}}_{p,m}
        \bigl(g^{\mathrm{PCC}}_{1,m}\bigr)^*\right]\right\}.
    \label{eq:pcc_gauge2}
\end{equation}
where $g^{\mathrm{PCC}}_{p,m}=h^{\mathrm{PCC}}_{p,m+G}\bigl(h^{\mathrm{PCC}}_{p,m}\bigr)^*$.
For nonzero gauge products and correct unwrapping, the noiseless model yields the same gauge-differential phase change as error-free first-period RVS. In fact, first-period referencing in~\eqref{eq:pcc_gauge2} retains the first gauge measurement's noise as a temporal phase offset. 

\section{Experimental Results and Discussion}\label{Exp}
Experiments use the coherent configuration in Fig.~\ref{fig1} over approximately \(7.1\)~km of fiber. A Precilasers FL-SF-1550-S narrow-linewidth laser provides the probe and unshifted LO. A Keysight M8190A arbitrary waveform generator (AWG) drives an ID PHOTONICS OMFTV2 IQ modulator. The probe and backscatter are optically filtered and amplified. The return is mixed with the LO and detected by a balanced photodetector (BPD), then digitized by a RIGOL DHO4404 digital storage oscilloscope (DSO) at \(1\)~GSa/s for offline processing. A piezoelectric transducer (PZT; Halliburton PZ1-SMF4-APC-E) located approximately \(5.1\)~km from the fiber input is driven at \(2\)~V to simulate a \(500\)-Hz sinusoidal vibration. 

We compare the proposed PCC in Section~\ref{Proposed} with first-period RVS~\cite{ChenOE2017} using the same \(T_{\mathrm b}=16~\mu\mathrm{s}\) LFM pulses at AWG output settings of \(441\) and \(331\)~mV$_{\mathrm{pp}}$. The sweep bandwidth is \(B=400\)~MHz, the heterodyne offset is \(f_{\mathrm{IF}}=240\)~MHz, and the pulse repetition interval is \(T_{\mathrm r}=120~\mu\mathrm{s}\). The proposed PCC combines the full-band replica and eight overlapping subbands with \(B_{\mathrm o}=200\)~MHz and \(\zeta=6/7\), whereas first-period RVS uses four nonoverlapping \(100\)-MHz subbands. Both methods use noise whitening by~\eqref{eq:whitenin} and the same gauge length of \(L_{\mathrm g}\approx3.06\)~m. 

Fig.~\ref{fig:recovered_phase} compares gauge-differential phase traces over all probing periods and local slow-time--distance maps. Under the reliability thresholds used for each method, the proposed PCC shows smaller background phase excursions and increases the vibration-to-background ratio (VBR) by \(2.76\) and \(2.46\)~dB at \(441\) and \(331\)~mV$_{\mathrm{pp}}$, respectively. Here, \(\mathrm{VBR}=20\log_{10}(\mu_{\mathrm v}/\mu_{\mathrm b})\), where \(\mu_{\mathrm v}\) and \(\mu_{\mathrm b}\) are the mean absolute gauge-differential phase over valid observations within the vibration window and over all remaining retained distance bins. This improvement is consistent with both more reliable coherent combining and a larger replica bank. The proposed PCC estimates weights across probing periods and combines the full-band replica with eight overlapping subbands. In contrast, first-period RVS aligns nonoverlapping subbands using a noisy first-period reference~\eqref{eq:reference_phase_error}. 

The local maps show periodic vibration-induced phase for both methods. Fig.~\ref{fig:recovered_phasetrace} further shows that the spatially gauge-differential phase within the vibration window closely follows a sine fit, with a clear \(500\)-Hz peak in the one-sided power spectral density (PSD). Thus, first-period RVS recovers the excited region despite background excursions that obscure its spatial contrast, while the proposed PCC improves background rejection. 
The proposed scheme combines \(B/2\) overlapping subbands with the full-band replica, whose nominal spatial-resolution scales are at least a factor of two smaller than that of the \(B/4\) RVS subbands. Note that the final spatial resolution also depends on the filter windows, combining weights, and gauge length. Processing nine replicas and calibrating their weights increases computational cost relative to the first-period RVS with four subbands.

\section{Conclusions}\label{Conclu}
This paper proposed a DAS receiver integrating frequency-domain replica reconstruction and noise-whitened PCC for Rayleigh-fading suppression. Experiments over approximately \(7.1\)~km of fiber recovered \(500\)-Hz vibrations and improved VBR by \(2.46\)--\(2.76\)~dB over first-period RVS under the applied reliability thresholds. Combining wider overlapping subbands with the full-band replica improves background rejection while easing the nominal spatial-resolution penalty, though it increases computational cost. This frequency-domain replica reconstruction uses a single probe return and can extend beyond LFM to other known probe waveforms. 


\begin{thebibliography}{99}
\bibitem{JoussetNC2018}
P.~Jousset \emph{et al.}, ``Dynamic strain determination using fibre-optic cables allows imaging of seismological and structural features,'' \emph{Nat. Commun.}, vol.~9, no.~1, Art.~no.~2509, 2018.

\bibitem{MartinsOL2013}
H.~F.~Martins, S.~Martin-Lopez, P.~Corredera, P.~Salgado, O.~Fraz{\~a}o, and M.~Gonz{\'a}lez-Herr{\'a}ez, ``Modulation instability-induced fading in phase-sensitive optical time-domain reflectometry,'' \emph{Opt. Lett.}, vol.~38, no.~6, pp.~872--874, 2013.

\bibitem{ZouOE2015}
W.~Zou, S.~Yang, X.~Long, and J.~Chen, ``Optical pulse compression reflectometry: proposal and proof-of-concept experiment,'' \emph{Opt. Express}, vol.~23, no.~1, pp.~512--522, 2015.

\bibitem{LuOL2017}
B.~Lu \emph{et al.}, ``High spatial resolution phase-sensitive optical time domain reflectometer with a frequency-swept pulse,'' \emph{Opt. Lett.}, vol.~42, no.~3, pp.~391--394, 2017.

\bibitem{LiokumovichJLT2015}
L.~B.~Liokumovich, \emph{et al.}, ``Fundamentals of optical fiber sensing schemes based on coherent optical time domain reflectometry: Signal model under static fiber conditions,'' \emph{J. Lightwave Technol.}, vol.~33, no.~17, pp.~3660--3671, 2015.

\bibitem{ZhouJLT2013}
J.~Zhou, Z.~Pan, Q.~Ye, H.~Cai, R.~Qu, and Z.~Fang, ``Characteristics and explanations of interference fading of a \(\phi\)-OTDR with a multi-frequency source,'' \emph{J. Lightwave Technol.}, vol.~31, no.~17, pp.~2947--2954, 2013.

\bibitem{GuPJ2020}
J.~Gu \emph{et al.}, ``High SNR \(\Phi\)-OTDR based on frequency and wavelength diversity with differential vector aggregation method,'' \emph{IEEE Photon. J.}, vol.~12, no.~6, Art.~no.~7103612, 2020.

\bibitem{ZhaoOE2021}
Z.~Zhao \emph{et al.}, ``Interference fading suppression in \(\phi\)-OTDR using space-division multiplexed probes,'' \emph{Opt. Express}, vol.~29, no.~10, pp.~15452--15462, 2021.

\bibitem{XiaoOE2024}
Y.~Xiao \emph{et al.}, ``Fading suppression and noise reduction of a DAS system integrated multi-core fiber,'' \emph{Opt. Express}, vol.~32, no.~15, pp.~26793--26807, 2024.

\bibitem{WuJLT2019}
Y.~Wu, Z.~Wang, J.~Xiong, J.~Jiang, S.~Lin, and Y.~Chen, ``Interference fading elimination with single rectangular pulse in \(\Phi\)-OTDR,'' \emph{J. Lightwave Technol.}, vol.~37, no.~13, pp.~3381--3387, 2019.

\bibitem{WuJLT2024}
Y.~Wu \emph{et al.}, ``Three-layer structure multiplexing fading elimination method in long-haul \(\Phi\)-OTDR,'' \emph{J. Lightwave Technol.}, vol.~42, no.~14, pp.~5017--5024, 2024.

\bibitem{ChenOE2017}
D.~Chen, Q.~Liu, and Z.~He, ``Phase-detection distributed fiber-optic vibration sensor without fading-noise based on time-gated digital OFDR,'' \emph{Opt. Express}, vol.~25, no.~7, pp.~8315--8325, 2017.

\bibitem{HuangDMTarXiv2026}
H.~Huang,  \emph{et al.}, ``A continuous payload-bearing discrete multitone modulation framework for fiber-optic integrated sensing and communication,'' \emph{arXiv preprint}, arXiv:2608.29020, 2026.

\bibitem{HuangCPOFDMarXiv2026}
H.~Huang \emph{et al.}, ``Cyclic-prefix OFDM probing for spatial-ISI-free distributed acoustic sensing via frequency-domain channel reconstruction,'' \emph{arXiv preprint}, arXiv:2606.19724v2, 2026.

\bibitem{LiuTWC2007}
T.-H.~Liu, ``Analysis of the principal component combining with weight estimation for DPSK signals,'' \emph{IEEE Trans. Wireless Commun.}, vol.~6, no.~3, pp.~822--826, 2007.

\end{thebibliography}
\end{document}